\documentstyle[12pt]{article}
\title{\Large \bf On  Galilean invariance and nonlinearity in
electrodynamics and quantum mechanics}
\author{ 
\bf     Gerald A. Goldin 
\\
\it   Departments of Mathematics and Physics, Rutgers University, \\
\it  Busch Campus, Piscataway, New Jersey 08854\\ 
e-mail: gagoldin@dimacs.rutgers.edu\\and\\
\bf  Vladimir Shtelen\\
\it   Department of Mathematics, Rutgers University, Busch Campus,\\
\it   Piscataway, New Jersey 08854\\ e-mail: shtelen@math.rutgers.edu
\\
}

\date{3 May 2000}
\begin{document}
\setcounter{section}{0}
\renewcommand{\theequation}{\arabic{section}.\arabic{equation}}
\renewcommand{\vec}{\bf}
\maketitle

\begin{abstract}
Recent experimental results on slow light
heighten interest in nonlinear
Maxwell theories. We obtain Galilei
covariant equations for electromagnetism 
by  allowing special nonlinearities
in the constitutive equations only,
keeping Maxwell's equations
unchanged. 
Combining these with linear or nonlinear
Schr\"odinger equations, e.g. as proposed by
Doebner and Goldin, yields a
Galilean quantum electrodynamics.
\end{abstract} 

\section{Introduction}
In this letter we propose
nonlinear constitutive equations
restricting the symmetry of Maxwell's equations to Galilean
symmetry. We also
show that these constitutive equations arise as the
formal nonrelativistic limit (taking the
speed of light $c\to \infty$) of the
nonlinear relativistic theory: no
additional hypotheses about relative
field strengths are needed.
Admitting such nonlinear constitutive equations
leads to possibilities that are entirely new.
Maxwell's equations in our
approach stay unchanged; it is the choice of the constitutive
equations alone  that makes the difference between relativistic and
nonrelativistic theories. We stress that the resulting
Galilean electrodynamics is essentially  nonlinear, although gauge invariant; linear
constitutive equations are indeed incompatible with Maxwell's equations 
and Galilean covariance.
Further, we extend the Galilean symmetry
to the minimally-coupled Schr\"odinger-Maxwell
theory, and to the coupled systems of Maxwell
and nonlinear Schr\"odinger equations as
proposed by Doebner and Goldin. Thus one
has a consistent and fully
Galilean covariant (but nonlinear) quantum
electrodynamics.

There is a commonly held ``folk belief''
among physicists that Maxwell's
equations, being Lorentz invariant,
are not consistent with Galilean symmetry. However
it is known (but not widely appreciated) that
Galilean covariance, like Lorentz covariance,
is a property of  all four of Maxwell's
equations for media in classical electrodynamics.
More than 25 years ago, Le Bellac
and L\'evy-Leblond emphasized this in  \cite{LBLL}, noting that the
clash between Maxwell's equations and Galilean relativity occurs only in
the constitutive equations.
Keeping the standard constitutive equations for the vacuum,
and introducing some additional conditions for the
electromagnetic fields, these authors derived
two distinct Galilean
limits of Maxwell's equations: one in which Faraday's law is lost
(called the electric limit); and another in which the
displacement current is zero, that violates the continuity
equation for charge and current densities
(called the magnetic limit).
Brown and Holland, in a recent discussion of these results,
observed as follows:
``It is noted that no fully Galilean-covariant theory
of a coupled Schr\"odinger-Maxwell system (where the density
and current of the Schr\"odinger field act as source of the
nonrelativistic Maxwell field) is possible.'' \cite{BH1999}

Dyson, in discussing an unpublished 1948 ``proof'' of Maxwell's
equations by Feynman, remarks, ``The proof begins with assumptions invariant under
Galilean transformations and ends with equations invariant under
Lorentz transformations. How could this have happened?''  \cite{D90}.
Dyson's paper immediately provoked a heated discussion in the literature;
see, for example \cite{com1, RH92, VF91}. In particular,
Vaidya and Farina \cite{VF91} ask in the title of their paper,
``Can Galilean mechanics and
full Maxwell equations coexist peacefully?'' and answer forcefully, ``No,
they cannot.''

 The fact that Maxwell's equations for media are both Lorentz and
Galilei invariant sheds  light on the mystery behind   Feynman's
insight. The conclusions of Brown and Holland, and Vaidya and Farina,
depend on the implicit assumption of
linear constitutive equations. In this letter
we take a completely different point of view.
We show how the choice between a Lorentz or Galilei
invariant theory can be made by introducing a special
class of nonlinearities in the
constitutive equations, {\it without modifying any of 
Maxwell's equations (or the continuity
equation).} Unlike Le Bellac
and L\'evy-Leblond we are not forced by constitutive equations
to choose between a privileged frame of reference (the ether) and
Einsteinian relativity.

Our interest in nonlinearity and Maxwell's equations
is heightened by recent experimental results in
quantum optics, including
optical squeezing and quantum non-demolition measurements,
that suggest the fundamental importance
of nonlinear constitutive equations in media.
In particular, we are motivated by
recent experimental results
demonstrating extremely slow light speed (as slow as
17 m/sec) in laser-dressed ultra-cold atomic media \cite{slowlight}
to look again at the question of Galilean-invariant electrodynamics
in the context of nonlinear, coupled Schr\"odinger-Maxwell
theories. The result is the framework for Galilean quantum
electrodynamics described here.
We would like to raise the question whether the constitutive equations
in our Galilean class could be a good mathematical model for some real
media, in some range of velocities and field strengths.

\section{Symmetry of Maxwell's  equations for media}
 
\setcounter{equation}{0}

A systematic investigation of symmetry of Maxwell's equations for
media 
was reported in Refs. \cite{FSS, FT}, where it was shown that these equations
are covariant under the inhomogeneous  group of general linear transformations
$GL(4,\bf {R})$, that includes both Lorentz and Galilei
transformations. General constitutive
equations restricting this  symmetry to
Poincar\'e symmetry were found.

It is
important for further analysis to use a system of units such as  the SI system, that avoids
incorporating  $c$ into the definition of the fundamental
fields. Without such a choice it is not possible to see how
nonrelativistic theory arises from a relativistic theory in the limit
 $c\to \infty$.
Maxwell's  equations for media written  in SI units  
have the form \cite{J99,B80,V93}:
$$
{\bf \nabla} \times {\vec E} =-\frac{\partial \bf B}{\partial t},$$ 
$${\bf \nabla} \cdot {\vec B} =0;$$
\begin{equation} {\vec \nabla} \times {\vec H} =\frac{\partial \bf D}{\partial t} + \vec j,\end{equation}
$$\vec \nabla \cdot \vec D =\rho,$$
where  $\vec E$ is the electric field, $\vec D$ is the electric displacement,
$\vec B$ is the magnetic induction, and $\vec H$ is the magnetic field; $\rho$
and $\vec j$ are  charge and current densities.
 The physically detectable fields are $ \vec E$ and $\vec B$, via the Lorentz force
on a charged particle  ${\vec F}=q({\vec E} +{\vec v }\times {\vec
B})$. The fields $\vec H$ and $\vec D$ may be regarded as  constructs used to
describe (via the constitutive equations)  how the directly observable fields
are  produced by charges and currents.

As  is well known, Maxwell's  equations are Lorentz covariant; in particular,
they are covariant under the space-time transformations  \cite{J99,B80,V93}\\ 
\begin{equation}
x'_{\parallel}=\gamma({\bf x} -{\bf v} t)_{\parallel}, \quad  x'_{\perp} =x_{\perp}, 
 \quad  t'=\gamma(t-\frac{{\bf v} \cdot {\bf x}}{c^2}),  
\end{equation}
$$ \gamma= \frac{1}{\sqrt{1-v^2/c^2}};$$
with the corresponding field transformations
$$
B'_{\parallel}= B_{\parallel}, \quad  B'_{\perp}=\gamma({\vec B}-\frac{1}{c^2}{\vec v}
\times {\vec E})_{\perp}, $$
$$E'_{\parallel}= E_{\parallel}, \ E'_{\perp}=\gamma({\vec E}+{\vec v}
\times {\vec B})_{\perp}, $$
\begin{equation}
H'_{\parallel}= H_{\parallel}, \ H'_{\perp}=\gamma({\vec H}-{\vec v}
\times \vec D)_{\perp},
\end{equation}
$$D'_{\parallel}= D_{\parallel}, \ D'_{\perp}=\gamma({\vec D}+\frac{1}{c^2}{\vec v}
\times {\vec H})_{\perp}; $$
and the transformations for current and charge densities:
\begin{equation}
j'_{\parallel}=\gamma({\vec j} -{\vec v} \rho)_{\parallel}, \quad  j'_{\perp} =j_{\perp}, 
 \quad  \rho'=\gamma(\rho -\frac{\bf v \cdot \bf j}{c^2}).
\end{equation}
It is appropriate to list here the Lorentz field invariants:
$$
I_1 ={\vec B}^2 -\frac{1}{c^2}{{\vec E}^2}, \quad I_2= \vec B\cdot\vec E;$$
\begin{equation}
I_3={\bf D}^2 -\frac{1}{c^2}{\vec H^2}, \quad  I_4={\bf H} \cdot \vec D;
\end{equation}
$$ I_5={\bf B} \cdot {\vec H} - {\bf E} \cdot {\vec D}, \quad  
I_6={\bf B }\cdot {\bf D} + \frac{1}{c^2} {\bf E} \cdot {\bf H}.$$

One can also verify that Maxwell's equations (2.1)
are  covariant 
with respect to the  Galilean transformations
$$t'=t,\quad \vec x'= \vec x -\vec v t;$$
$$\vec E'= \vec E + \vec v \times \vec B, \quad \vec B'= \vec B;$$
\begin{equation}
\vec H'= \vec H - \vec v \times \vec D, \quad \vec D'= \vec D ;
\end{equation}
$$\vec j'=\vec j - \rho\vec v, \quad \rho '= \rho  ,$$
which arise  as the  $c\to\infty$ limit of the Lorentz transformations
(2.2) - (2.4). Both  Galilean and
Lorentz symmetries belong to the class  of general linear transformations $GL(4,\bf {R})$
admitted by Maxwell's equations (2.1). This, in turn, is a
consequence of the incompleteness of the system (2.1): there are  8 equations for 12 unknown functions.
The system  must  be completed by the  constitutive
equations, which are functional relations between vectors $\vec D,
\   \vec E, \  \vec B, $ and  $ \vec H.$  

A particular choice of the
constitutive equations can reduce the symmetry of the system (2.1) to
Lorentz  or to Galilei.   General constitutive
equations that restrict the  $GL(4,\bf {R})$ symmetry  of Maxwell's equations
to Lorentz (Poincar\'e) are reported in Refs.  \cite{FSS, FT}. In SI units,
they have the  form 
 \begin{equation}
{\bf D} = M{\vec B} + \frac{1}{c^2}N{\bf E}, \quad {\vec H}=N{\vec B} - M{\vec E},
\end{equation}
where $M$ and $N$ are  arbitrary scalar functions of the Lorentz  invariants
$I_1, \  I_2$  given in (2.5).
Equivalently we may write
 \begin{equation}
{\vec B} =R{\vec D} + \frac{1}{c^2}Q{\vec H}, \quad {\vec E}=Q{\vec D} - R{\vec H}, 
\end{equation}
where $Q$ and $R$ are arbitrary functions of the Lorentz invariants
$I_3$, $I_4$ from (2.5).

Let us note that some well-known  nonlinear
theories like Born-Infeld electrodynamics, or Euler-Kockel
electrodynamics that takes into account quantum-mechanical nonlinear
effects (see, for example, \cite{J99} and references therein), correspond to  
particular choices
of $M$ and $N$ in (2.7).

The constitutive equations 
 that reduce the $GL(4,\bf {R})$ symmetry group  of (2.1) to the
Galilei group are
\begin{equation}
{\bf D} = \hat{M} {\bf B}, \quad {\bf H}=\hat{N} {\bf B} - \hat{M}\bf E,
\end{equation}
 or, equivalently
 \begin{equation}
{\bf B} = \hat{R}{\vec D}, \quad {\vec E}=\hat{Q}{\vec D} - \hat{R}{\vec H}, 
\end{equation}
 where $\hat{M}$ and  $\hat{N}$, $ \hat{Q}$ and $\hat{R}$ are arbitrary functions of 
Galilean invariants. To demonstrate this, we first use (2.6) to see
that  the general  constitutive equations
$\vec D = \vec f(\vec B,  \vec E ),$
$\vec H = \vec g(\vec B,    \vec E)$ must
take  the form
$
{\vec D} =\hat{M} {\vec B}, \quad {\vec H} = \hat{N} {\vec B} + \hat{N_1} {\vec E},
$
where $\hat{M},  \hat{N},  \hat{N_1}$ are some scalar  functions of Galilean
invariants. Then  substitution of the latter equation into (2.6) results in the condition
$\hat{N_1}=-\hat{M}$. Finally, one shows
 that the Galilean field invariants are:
$$
\hat{I_1} ={\vec B}^2 , \quad \hat{I_2}= \vec B \cdot \vec E;$$
\begin{equation}
\hat{I_3}={\vec D}^2 , \quad  \hat{I_4}=\vec H\cdot\vec D;
\end{equation}
$$\hat{I_5}={\vec B} \cdot {\vec H} - {\vec E} \cdot {\vec D}, \quad  
\hat{I_6}=\vec B\cdot\vec D .$$ 
One can, if one wishes,  restrict
oneself to  constitutive equations in explicit (i.e., non-implicit) form, so that 
$\hat{M}$ and  $\hat{N}$ depend on  $\hat{I_1}, \hat{I_2}$, and
$\hat{Q}$ and $\hat{R}$ depend on  $\hat{I_3}, \hat{I_4}$.

 Note that the  constitutive equations (2.9),  (2.10) and the
Galilean field invariants (2.11) are respectively the formal limits as 
$c\to\infty$ of the Lorentz invariant relations (2.7), (2.8), and (2.5).

 One obtains the standard Maxwell equations for the vacuum
(where $c^{-2}= \epsilon_0\mu_0$)
by choosing  $N=1/\mu_0,$  $ M=0$  in (2.7).
But in the case of Galilean constitutive equations,  letting $M$ be a
constant  is not compatible
 with a  nonzero charge density  $\rho$. 
Therefore,  we
conclude that a  consistent Galilean electrodynamics is {\it
essentially nonlinear}. 

It is worth  remarking  for  the case of  relativistic constitutive equations
(2.7), that letting  $M=\lambda =$ constant,  so that  
$$
{\bf D} = \lambda{\vec B} + \frac{1}{c^2}N{\bf E}, \quad 
{\vec H}=N{\vec B} -\lambda {\vec E},
$$
results in  equations for the  observable fields $ {\bf B}$  and  ${\bf E}$ that
are independent of the magnitude of $\lambda$. In other words,
constant terms
 $\lambda$ in the above constitutive equations have no effect on
 observable fields, and therefore, without loss of generality one can
set  $\lambda=0$. In general, adding a constant to $M$ in (2.7) has no effect
on the  observable fields ${\bf E}$ and  ${\bf B}$. This is also true
for $\hat{M}$ in (2.9), which is why nontrivial Galilean constitutive
equations are essentially nonlinear.

The (linear)  Galilean  electrodynamics discussed in the
literature  is 
associated with   the pre-Maxwell electrodynamics,  with no displacement
current \cite{LBLL, BH1999, J99, V93}, it
is also called the ``magnetic  limit''  of Maxwell's equations,
and is the only one of the two limits discussed in [1] that is
compatible with Schr\"odinger quantum mechanics. These equations in SI units  have the form:
$$
{\bf \nabla} \times {\vec E} =-\frac{\partial {\vec B}}{\partial t},$$ 
$$ {\bf \nabla} \cdot {\vec B} =0;$$
\begin{equation} {\vec \nabla} \times {\vec B} =\mu_0 {\vec j},\end{equation}
$${\bf \nabla} \cdot {\vec E} ={\epsilon_0}^{-1}\rho.$$
Let us note, that the approach used in \cite{LBLL} to  derive  (2.12)
from  (2.1) actually embodies contradictory assumptions,
  moving back and forth between the two symmetries. First, the
 constitutive equations for the vacuum ${\vec D}
={\epsilon_0} {\vec E}, \quad  {\vec B}=\mu_0 {\vec H}$ are imposed, 
immediately breaking  the Galilean symmetry of Maxwell's equations (2.1). 
Then additional, non-invariant  conditions   about
relative field strengths  are assumed to justify the limit $1/c^2 \to 0$ in Maxwell's
equations,  and  remove the displacement current. At the same
time   both
${\epsilon_0}$ and $\mu_0$ are kept non-zero, even though their
product $1/c^2$ has been taken to zero.
Since there is no displacement current in (2.12), the continuity
equation does not hold. The Galilean transformations for  $\vec E$ and
$\vec B$ in (2.12) are as in
(2.6), while the corresponding transformations for $\rho$ and $\vec j$
are
\begin{equation}
{\vec j'}={\vec j} ,  \quad \rho '= \rho -{\epsilon_0}\mu_0 \vec v \cdot\vec j.
\end{equation}

In order to explore the consistency of the Galilean electrodynamics
(2.1), (2.9), we have investigated two situations: first, the case $\hat{M}=0$,
  $\hat{N}=\vec E\cdot\vec B$ (which requires $\rho=0$); secondly,  
$\hat{M}=\vec E\cdot\vec B$,  $\hat{N}=1/\mu_0$.
We were able to obtain some particular solutions, and
this is a  subject of our ongoing research. Let us note here that
using the Galilei invariance of the system, one can construct (new)
 traveling wave solutions from  (old) solutions
 by means of the following formulas:
$${\vec E}_{new}(x)= {\vec E}_{old}(x') - {\vec v} \times {\vec B}_{old}(x'),
\quad {\vec B}_{new}(x)= {\vec B}_{old}(x');$$
\begin{equation}
{\vec H}_{new}(x)= {\vec H}_{old}(x') + {\vec v} \times {\vec
D}_{old}(x'), \quad 
{\vec D}_{new}(x)= {\vec D}_{old}(x') ;
\end{equation}
$${\vec j}_{new}(x)={\vec j}_{old}(x') + {\vec v} \rho_{old}(x'), \quad \rho _{new}(x)= \rho_{old}(x')  ,$$
where $ x'=({\bf x'}, t')=({\vec x} - {\vec v} t, t)$.




\section{A framework for a Galilei covariant quantum electrodynamics}
 \setcounter{equation}{0}

As  is well known, any solution for  vectors $\vec E$  and $\vec B$ of Maxwell's equations (2.1), 
 can be represented in terms of potentials
($\Phi, \  \vec A$)   
$$
{\vec B}={\vec \nabla} \times {\vec A} ,
$$
\begin{equation}
{\vec E} = -\frac{\partial {\vec A}}{\partial t}-{\vec \nabla}
\Phi .
\end{equation}
for some   $\Phi$ and $ \vec A$. The choice of $\Phi$ and $ \vec A$ is
not unique:  new potentials
$
{\vec A}'={\vec A} + {\bf \nabla} \Lambda,$
$\Phi'=\Phi-\partial \Lambda/{\partial t},$
containing an arbitrary function $\Lambda$,  result in  the same $\vec E$  and
$\vec B$  (gauge invariance). 
The standard procedure of introducing
electromagnetic interaction in quantum mechanics  is  minimal
coupling, which is consistent  
 with Galilean covariance
 of the free field equations.
    One   obtains the  coupled Schr\"odinger-Maxwell equation
\begin{equation}
    i\hbar\frac{\partial\psi}{\partial t} = \frac{1}{2m}(-i\hbar{\vec
\nabla} - e{\vec A})^2\psi + e\Phi\psi , 
\end{equation}
where the electromagnetic fields $\vec E, \vec B$ are obtained from
$\Phi, \vec A$ via (3.1), and are governed by (2.1). Gauge invariant  current and charge
densities  entering (2.1) are given by
\begin{equation}
\rho=\bar\psi\psi, \quad {\bf J^{gi}}=\frac{\hbar}{2im}[\bar\psi{\bf
\nabla}\psi-({\bf \nabla}\bar\psi)\psi]-
\frac{e}{m}\rho\vec A.
\end{equation}
Equation (3.2) is Galilean invariant, with corresponding
transformations  given in (2.6),  and the transformations for  vector-potential $(\Phi, \vec A)$
given by
\begin{equation}
{\bf A}'= {\bf A}, \quad   \Phi'= \Phi-\vec v \cdot \vec A.
\end{equation}
The resulting fully Galilean covariant  quantum electrodynamics is embodied in equations
(2.1), (2.9) (with some concrete choices of $\hat M$ and $\hat N)$, (3.1), (3.2), (3.3). 

It is natural to ask about the coupling of nonlinear Maxwell theory
with nonlinear Galilean invariant Schr\"odinger mechanics.
A possible framework for such a quantum mechanics is given by the
family of  nonlinear equations proposed by Doebner and Goldin  \cite{DG96}.
Nonlinear terms of the form
\begin{equation}
\frac{i\hbar D}{2}\frac{\triangle \rho}{\rho}\psi + \hbar
D'[c_1\frac{{\bf \nabla}\cdot \hat {\bf j}}{\rho} + 
c_2\frac{\triangle \rho}{\rho} +
c_3\frac{{\hat {\bf j}}^2}{\rho^2} +
c_4\frac{{\hat {\bf j}} \cdot {\bf \nabla}\rho}{\rho^2} +
c_5\frac{{({\bf \nabla} \rho)}^2}{\rho^2}]\psi
\end{equation}
are added to the right side of Eq.(3.2), where 
$\hat{ \vec
j}=(1/2i)[\bar\psi\vec\nabla\psi-(\vec\nabla\bar\psi)\psi]$, 
and $D, D'$ are diffusion coefficients. The Galilean-invariant 
subfamily of quantum theories is defined by 
$c_1+c_4=c_3=0$. We are proposing consideration of the coupled system
of Doebner-Goldin equations and nonlinear Galilean invariant Maxwell
equations, as a consistent Galilean quantum electrodynamics. In this
system, the quantities $\rho, {\bf E}$,  ${\bf B}$,  and
\begin{equation}
{\bf J^{gi}}=\frac{\hbar}{2im}[\bar\psi{\bf
\nabla}\psi-({\bf \nabla}\bar\psi)\psi]-D{\bf \nabla}\rho - 
\frac{e}{m}\rho\vec A.
\end{equation}
are gauge invariant. Eq.(3.6) gives the current to be entered in
Maxwell's equations. Doebner
 and Goldin further  generalize the notion of
gauge transformation to allow nonlinear gauge  transformations,
writing more general formulas for the gauge invariants
$\rho, {\bf J^{gi}}, {\bf E}$, and ${\bf B}$ \cite{G97}.
Recently Galilean invariant Doebner-Goldin equations  were applied
to describe  the dynamics of matrix D-branes \cite{MS}.

In conclusion, let us remark  that one can  extend our  approach further to
include in the theory Galilei invariant spinor equations (see, for example, \cite{FSS}). Another
possible generalization of the quantum electrodynamics described here is
to non-Abelian Yang-Mills fields.


\begin{thebibliography}{}
%
\bibitem{LBLL} M. Le Bellac and J.-M. Levy-Leblond,
{Galilean electromagnetism},
Nuovo~Cim.~{\bf 14B} (1973) 217.
%
\bibitem{BH1999}
H.~R.~Brown, P.~R.~Holland,
{The Galilean covariance of
quantum mechanics in the case of external fields},
Am.~J.~Phys.~{\bf 67} (1999) 204.
%
\bibitem{D90} F.~J.~Dyson,
{Feynman's proof of the Maxwell equations}, 
Am.~J.~Phys~{\bf 58} (1990) 209.
%
\bibitem{com1} N.~Dombey,
{Comment on ``Feynman's proof of the Maxwell
equations'', by F.J. Dyson},
Am.~J.~Phys~{\bf 59} (1991) 85.
%
\bibitem{RH92} R.~J. Hughes,
{On Feynman's proof of the Maxwell
equations},
Am.~J.~Phys~{\bf 60} (1991) 301.
%
\bibitem{VF91} A.~Vaidya and C.~Farina,
{Can Galilean mechanics and
full Maxwell equations coexist peacefully?}, 
Phys.~Lett.~A~{\bf 153} (1991) 265.
%
\bibitem{slowlight}
J.~Marangos,
{Slow light in cool atoms}, 
Nature~{\bf 397} (18~February 1999), 559.
%
\bibitem{FSS} W.~I.~Fushchich, V.~M.~Shtelen, and N.~I.~Serov,
{\it Symmetry
Analysis and Exact Solutions of Equations of Nonlinear Mathematical
Physics,\/}
Kluwer Acad.~Publ., Dordrecht (1993).
%
\bibitem{FT} W.~Fushchich and I.~Tsifra,
{On symmetry of nonlinear
equations of electrodynamics}, 
Teor.~Mat.~Fizika~{\bf 64} (1985) 41.
%
\bibitem{J99}
J.~D.~Jackson,
{\it Classical Electrodynamics,\/}
third edition,
J.~Wiley~\& Sons, Inc., New York (1999).
%
\bibitem{B80} A.~O.~Barut,
{\it Electrodynamics and Classical Theory of
Fields and Particles,\/}
Dover Publ., New York (1980).
%
\bibitem{V93} J.~Vanderlinde,
{\it Classical Electromagnetic Theory,\/}
Wiley, New York (1993).
%
\bibitem{DG96}H.-D.~Doebner and G.~A.~Goldin,
{Introducing nonlinear gauge
transformations in
a family of nonlinear Schr\"odinger equations}, 
Phys.~Rev.~A~{\bf 54} (1996) 3764.
%
\bibitem{G97} G.~A.~Goldin,
{Gauge transformations for a family of
nonlinear Schr\"odinger  equations}, 
J.~Nonl.~Math.~Phys.~{\bf 4} (1997) 6.
%
\bibitem{DGN99} H.-D.~Doebner,
G.~A.~Goldin, and P.~Nattermann,
{Gauge transformations in quantum mechanics
and the unification of nonlinear Schr\"odinger
equations},  
J.~Math.~Phys.~{\bf 40} (1999) 49.
%
\bibitem{MS} N.~E.~Mavromatos and R.~S.~Szabo,
{Nonlinear
Schr\"odinger dynamics of matrix D-branes}, 
Preprint (September 1999), OUTP-99-47P,
CERN-TH/99-256, NBI-HE-99-34, hep-th/9909129.


\end{thebibliography}
\end{document}